\documentclass[12pt, onecolumn,draftclsnofoot,journal]{IEEEtran}
\usepackage{epsfig,cite}
\usepackage{color}
\usepackage{upgreek}
\usepackage{amssymb, amsmath}
\usepackage{algorithm} 
\usepackage{algorithmic}
\usepackage{subfig}
\usepackage{stfloats}

\IEEEoverridecommandlockouts

\begin{document}


\newcounter{MYtempeqncnt}

\title{Design and Optimizing of On-Chip Kinesin Substrates for Molecular Communication}

\author{
Nariman Farsad, \IEEEmembership{Student Member, IEEE,} 
Andrew W. Eckford, \IEEEmembership{Member, IEEE,}\\ 
and Satoshi Hiyama, \IEEEmembership{Member, IEEE,}%
%
%
%
\thanks{This work as been presented in part at 2012 IEEE NANO conference.} 
\thanks{Nariman Farsad and Andrew W. Eckford are with the Department of Electrical Engineering and Computer Science, York University, 4700 Keele Street, Toronto, Ontario, Canada M3J 1P3. Emails: nariman@cse.yorku.ca, aeckford@yorku.ca}%
\thanks{Satoshi Hiyama is with Research Laboratories, NTT DOCOMO Inc., Yokosuka, Kanagawa, Japan. Email: hiyamas@nttdocomo.com}%
%
%
%
}

\maketitle


\begin{abstract}
Lab-on-chip devices and point-of-care diagnostic chip devices are composed of many different components such as nanosensors that must be able to communicate with other components within the device. Molecular communication is a promising solution for on-chip communication. In particular, kinesin driven microtubule (MT) motility is an effective means of transferring information particles from  one component to another. However, finding an optimal shape for these channels can be challenging. In this paper we derive a mathematical optimization model that can be used to find the optimal channel shape and dimensions for any transmission period. We derive three specific models for the rectangular channels, regular polygonal channels, and regular polygonal ring channels. We show that the optimal channel shapes are the square-shaped channel for the rectangular channel, and circular-shaped channel for the other classes of shapes. Finally, we show that among all 2 dimensional shapes the optimal design choice that maximizes information rate is the circular-shaped channel.        
\end{abstract}


\section{Introduction}
\label{sec:intro}

With the advancements in the field of nanotechnology, applications such as lab-on-chip devices \cite{daw06}, and point-of-care diagnostic chips \cite{yag06} are becoming a reality. In most of these applications communication between different components in each device, or between a biological entity and a component in the device, such as a nanosensor, is required. Therefore, one of the obstacles that must be overcome before many of these applications can be fully realized, is devising a communication system for small scales \cite{aky2008, bush-book}.

Inspired by nature, one of the most promising solutions is {\em molecular communication} \cite{hiy05, hiy10NanoCom}, where molecules released by a transmitter device propagate in a fluidic environment and transfer information to a receiver device. Information can be conveyed by encoding messages into the release timing \cite{eck09}, number \cite{far11NanoCom,far12NanoBio}, concentration\cite{mah10}, or identities of particles\cite{cob10}. Different types of propagation are possible for transporting information particles in on-chip molecular communication such as: diffusion \cite{nak08,pie10, pie13}, diffusion with flow \cite{sri12}, active transport using molecular motors and cytoskeletal filaments \cite{hiy10LabChip}, bacterial assisted propagation \cite{gre10, cob10, lio12}, and kinesin molecular motors moving over immobilized microtubule (MT) tracks \cite{moo09b,eno11}.

Molecular communication is biocompatible, and can be very energy efficient. Because of these properties, molecular communication has attracted a lot of attention in recent years \cite{nak12}. It has many potential applications in biomedical engineering from lab-on-chip devices \cite{bri13} and point-of-care diagnostic devices \cite{all11} to targeted drug delivery \cite{cha13}. However, most previous works have considered diffusion based molecular communication, which can be very slow for on-chip applications. Moreover, in \cite{far12NanoBio} it was shown that active transport using MT filaments gliding over kinesin covered substrate can generate higher information rates compared to flow based propagation over larger separations distances. Other types of propagations are either difficult to implement or have not been implemented yet. Moreover, in \cite{kim07, duj08,kim13} it is shown that electrical currents can be used to control the speed and direction of the MTs. This makes kinesin-driven MT motility a suitable form of propagation for on-chip molecular communication and transport, with potential applications in lab-on-chip devices, point-of-care diagnostic devices, organ-on-a-chip devices, and microfluid devices.

For the kinesin-microtubule based molecular communication to be fully realized, guide posts for design and implementation are required. For example, in microfluidics it is well known that different channel designs could be used to implement different functionalities in droplet microfluidics \cite{teh08}. However, finding the optimal design for these channels can be difficult because of the system complexities such as the random motion of the MTs. Moreover, experimental trial-and-error-based design is very laborious and time-consuming. Therefore, simulation environments based on experimentally validated models such as \cite{nit06,nit10} can be a very effective tool for providing design guide posts. For example in our previous works \cite{far11, far11NanoCom}, optimization of the shape of the transmission zone (i.e. location placement of transmitters) is considered using computer simulations, and it was shown that optimal transmission zone is along of the walls of the channel. This is because usually MTs follow the channel walls \cite{cle03}. Moreover, in \cite{far12Mona,far12NANO} we have shown that the shape of the channel can have a huge effect on information transmission rate, and we provide basic guidepost for designing the shape of the channel. 

In this work, we extend the preliminary results presented in our previous works and derive an optimization model that could be used to find the optimal channel shape and dimensions. Our results can be used in two ways. First, in on-chip transport applications, our results could be used to maximize the transport rate. On-chip transport is a very important process in lab-on-chip devices. For example, in \cite{ste14}, kinesin-microtubule transport scheme is used for high-throughput molecular transport and assembly. Our results can also be used in on-chip molecular communication applications to maximize the channel capacity of the system. We show that using the optimized channel shape can increase the capacity by more than 1.5 bits per channel use and in some cases by more than 2.5 bits per channel use. This can be a significant gain in sequential data transmission.

To setup our optimization formula, first we derive a complete model relating the number of MT trips during a single transmission period and the shape of the channel. This is our first major contribution.  We then use this model to find the optimal shape and dimensions for rectangular channels, regular polygonal channels, and regular polygonal ring-shaped channels. This is our second major contribution. Finally, we show that for all possible two dimensional channel shapes, the optimal channel is the circular-shaped channel. Moreover, for a particular transmission rate we show that we could use our optimization model to find the optimal radius for this circle. This is our third major contribution. We verify all our results using computer simulations.

The rest of this paper is organized as follows. In Section \ref{sec:ATMC} an overview of the kinesin-microtubule based molecular communication system is presented. A general information transmission model is presented in Section \ref{sec:chanCap}. In Section \ref{sec:model} we derive the channel shape optimization model, and in Section \ref{sec:OptiShapAna} we use the derived model to find the optimal channel design strategies. The results are then compared with computer simulations in Section \ref{sec:results}. Concluding remarks are given in Section \ref{sec:conclusion}.

\section{Kinesin-Microtubule based Molecular Communication Channels}
\label{sec:ATMC}

\begin{figure}
	\begin{center}
	\includegraphics[width=5in]{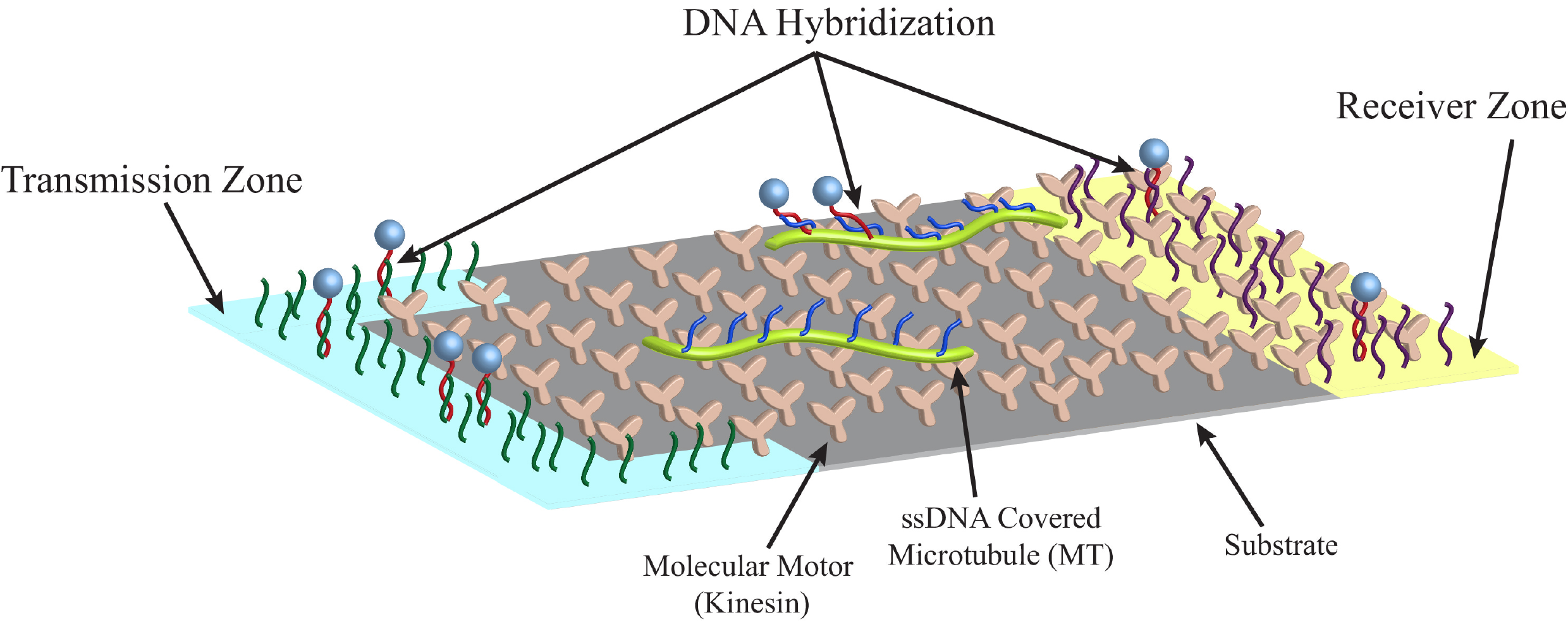}
	\end{center}
	\caption{\label{fig:MolComEnv} Depiction of the communication environment.}
\end{figure}
For our model we assume that an area in the kinesin covered channel is designated as the {\em transmission zone}, where the transmitter and the information particles released by the transmitter reside. Similarly, an area in the channel is designated as the {\em receiver zone}, where the receiver and its receptors are placed. Generally, we assume the transmission zone and the receiver zone are located at the opposite ends of the channel. Moreover, the information particles released by the transmitter remain stationary at the transmission zone until they are picked up by MT filaments gliding over kinesin covered substrate. The MT filaments then carry these information particles to the receiver zone, where they are unloaded and delivered to the receiver. The information can be encoded in the type, number, release timing, and concentration of the particles released and delivered. Moreover, the information particles can be encapsulated inside liposomes to protect them from denaturation (e.g., molecular deformation and cleavage caused by enzymatic attacks or changes in pH of the outer aqueous phase) in the propagation environment. 
 
For the loading and unloading information particles, we assume deoxyribonucleic acid (DNA) hybridization bonds are employed as shown in \cite{hiy09}. In this scheme, information particles, MT filaments, transmission zone, and the receiver zones are covered with single stranded DNAs (ssDNA). When information particles are released by the transmitter, they anchor themselves to the transmission zone through DNA hybridization bonds until an MT filament passes at the vicinity of the anchored particle. At this point the particle is loaded on to the passing MT again through DNA hybridization. When a loaded MT enters this receiver zone, the loaded particles are unloaded using yet another DNA hybridization bond. This process is summarized in Figure \ref{fig:MolComEnv} and the reader is referred to \cite{hiy09} for detailed explanation and laboratory demonstration of this technique.

\subsection{Shape of the Channel}
\label{sec:ChanShape}

\begin{figure}
	\begin{center}
	\includegraphics[width=5in]{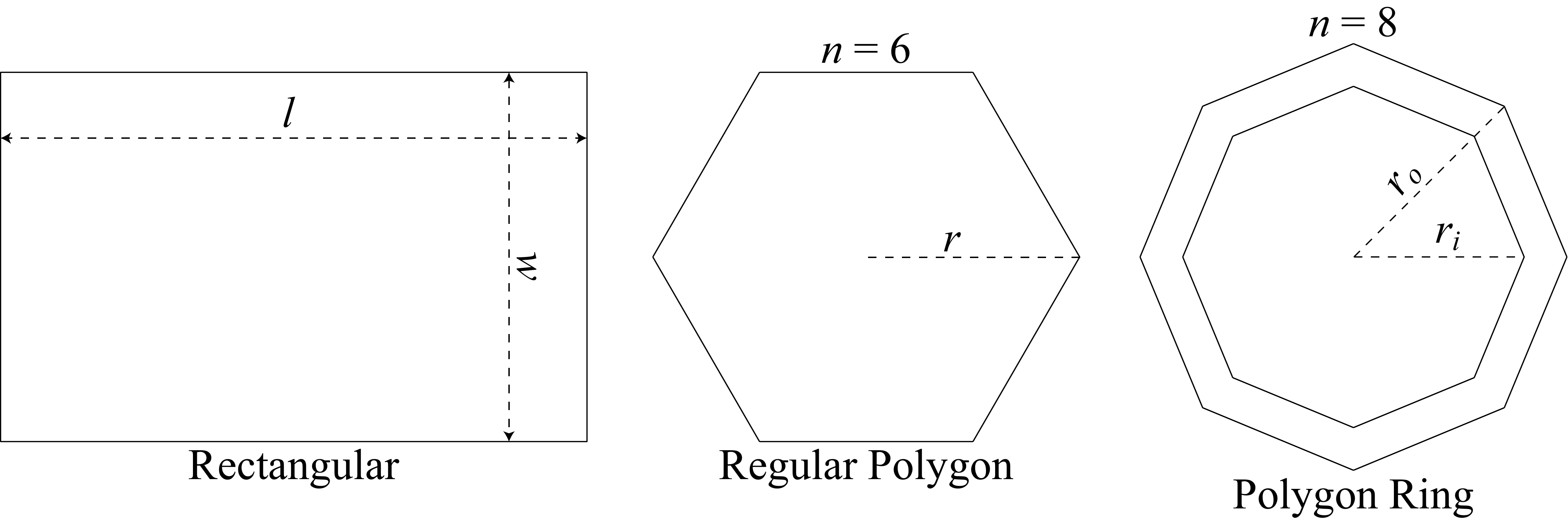}
	\end{center}
	\caption{\label{fig:ShapeClasses} Different shape classes and their parameters.}
\end{figure}

In this paper we assume the channel can be formed to have a wide variety of different shapes, and we find the optimal shape that maximizes information transmission rate among all possible shapes. In particular, we assume that channel shape can belong to any of the following three classes of shapes: rectangular, regular polygon, and regular polygon ring. Regular polygons, are equiangular (all angles are equal in measure) and equilateral (all sides have the same length), and they include a large class of geometric shapes, ranging from equilateral triangles, squares, pentagons, and hexagons, all the way up to a circle as number of sides approaches infinity. Figure \ref{fig:ShapeClasses} shows an example from each of the three different shape classes. In practice, these channels, as well as channel having any other shape can be created using similar procedures used in \cite{hiy09,hiy10LabChip,nit10}. 

Depending on the class of the channel shape, different parameters can be used to further adjust the shape within the shape class. For rectangular channels two parameters can be used to adjust the dimensions and the shape: width $w$ and length $l$. When both the width and the length are equal in value the channel would be a square channel. Similarly, for regular polygons two parameters can be used to further adjust the shape: the number of sides $n$, and the radius of the circumscribed circle $r$ (this radius can also be defined as the distance from the center point of the regular polygon to one of the edges of the polygon). For ring-shaped channels three parameters are used to define the shape: the number of sides $n$, the radius of the inner circumscribed circle $r_i$, and the radius of the outer circumscribed circle $r_o$. Figure \ref{fig:ShapeClasses} summarizes all these parameters for different classes of shapes.

Regardless of the shape of the channel, without loss of generality, we assume the transmission zone is always on the left side of the channel and the receiver zone is along the right side of the channel. In \cite{far11NanoCom}, it was shown that the optimal transmission zone is along the walls of the channel since MTs mostly glide very close to channel walls. This positioning increases the chance of an MT picking up an information particle during its trips. Therefore, in this work we always assume the transmission zone is along the walls of the channel regardless of the channel shape. Moreover, the minimum separation distance between the start of the transmission zone and the receiver zone are always the same in our numerical evaluations of Section \ref{sec:results} regardless of the shape of the channel.

\subsection{Simulating the Channel}
\label{sec:ChanSim}

In \cite{hiy09, hiy10LabChip} the authors show that creating kinesin-microtubule based channels described in the previous section is possible in wet labs. However, studying the resulting molecular communication channel, and solving design and optimization problems using laboratory experimentation is extremely difficult because these experiment are time consuming, laborious, and expensive. To overcome this issue, computer simulations have been used in previous works \cite{nit10, far11NanoCom, far12NanoBio}. Similarly, in this work we use Monte-Carlo techniques proposed in \cite{nit06} to simulate the motion of the MTs over a kinesin covered substrate. The equations for the motion of the MT, which were developed and verified experimentally in \cite{nit06}, are given below. 

The motion of the MT is largely regular, although the effects of Brownian motion cause random fluctuations. Since the MTs move only in the $x$--$y$ directions, and do not move in the $z$ direction (along the height of the channel), we consider a two-dimensional simulation of MTs for $\Delta t$ time intervals. Given some initial position $(x_0, y_0)$ at time $t = 0$, for any integer $k > 0$, the motion of the MT is given by the sequence of coordinates $(x_i, y_i)$ for $i=1, 2, \ldots, k$. Each coordinate $(x_i, y_i)$ represents the position of the MT's head at the end of the time $t = i \Delta t$, where
\begin{eqnarray}
	\label{eqn:MTXCoord}
	x_i & = & x_{i-1}  + \Delta r \cos \theta_i, \\
	\label{eqn:MTYCoord}
	y_i & = & y_{i-1}  + \Delta r \sin \theta_i . 
\end{eqnarray}
In this case, the step size $\Delta r$ at each step is an independent and identically distributed (i.i.d.) Gaussian random variable with mean and variance
\begin{eqnarray}
	E[\Delta r] & = & v_{\mathrm{avg}} \Delta t, \\
	\mathrm{Var}[\Delta r] & = & 2 D \Delta t,
\end{eqnarray}
where $v_{\mathrm{avg}}$ is the average velocity of the MT, and $D$ is the MT's diffusion coefficient. The angle $\theta_i$ is no longer independent from step to step: instead,
for some step-to-step angular change $\Delta \theta$, we have that
\begin{equation}
	\theta_i = \Delta \theta + \theta_{i-1} .
\end{equation}
Now, for each step, $\Delta \theta$ is an i.i.d. Gaussian-distributed random variable with mean and variance
\begin{eqnarray}
	E[\Delta \theta] & = & 0, \\
	\mathrm{Var}[\Delta \theta] & = & \frac{v_{\mathrm{avg}}\Delta t}{L_p} ,
\end{eqnarray}
where $L_p$ is the persistence length of the MT's trajectory. Following \cite{nit06}, in case of a collision with a boundary, we assume that the MT sets $\theta_i$ so as to {\em follow the boundary}. Finally, our simulator can model any polygon-shaped channel.

We also use the grid loading mechanism proposed in \cite{far11NanoCom}, to simulate the information particles' loading and unloading. For loading an information particle, the MT filament must drive close to the anchored particle. Therefore, we divide the transmission zone into a square grid, where the length of each square in the grid is the same as the diameter of the particles. We then distribute particles randomly and uniformly among the squares in the grid. In general, we assume that the MTs can load multiple particles, which we know to be possible based on lab experiments; thus, if an MT enters a square which is occupied by a particle, and it has an empty loading slot available, we assume the MT loads that particle. For unloading, we assume all the loaded particles are unloaded as soon as an MT enters the receiver zone.
\begin{figure}
	\begin{center}
	\includegraphics[width=3.4in]{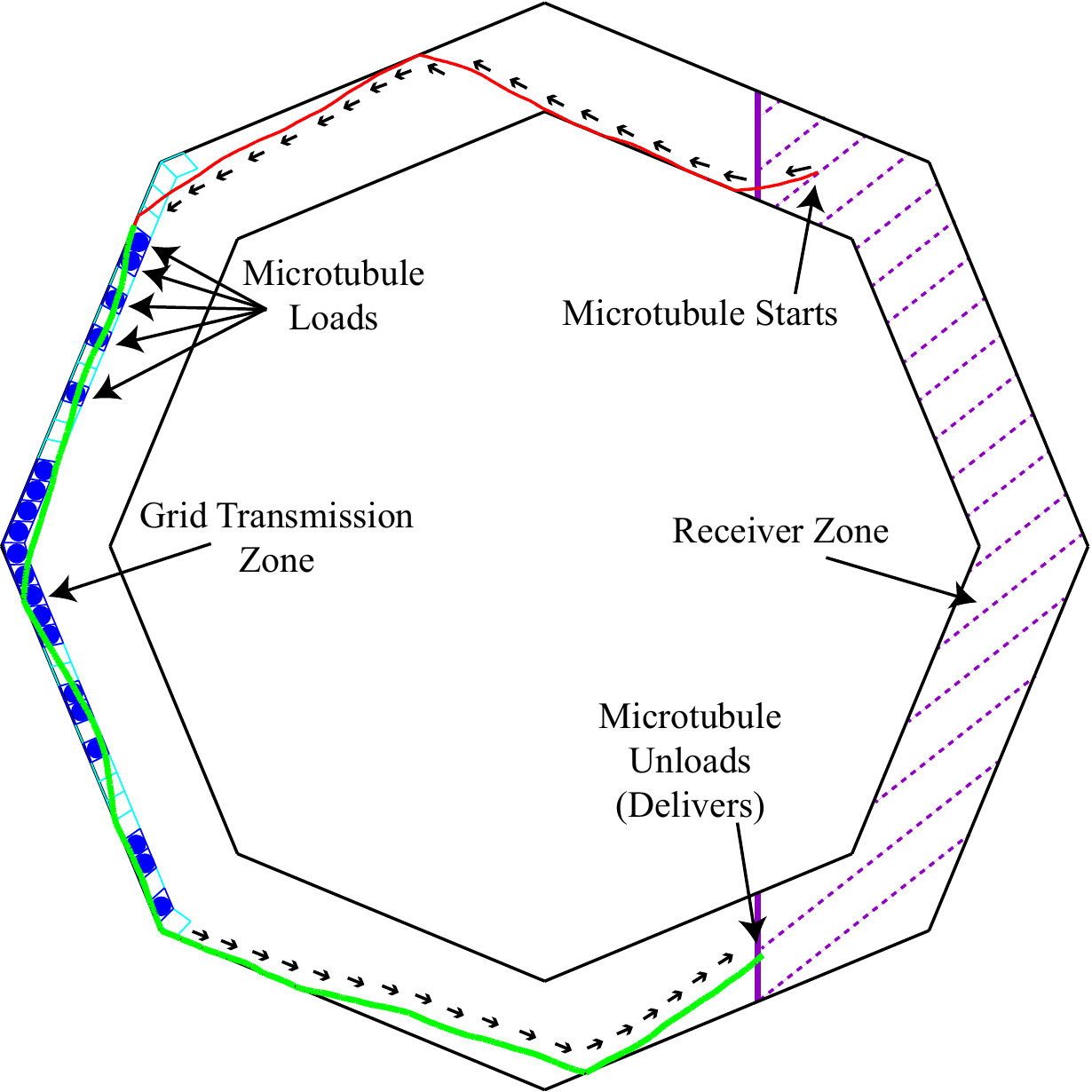}
	\end{center}
	\caption{\label{fig:SimEnv} An example of the simulation environment.}
\end{figure}

Figure \ref{fig:SimEnv} shows a sample simulation. In this figure, the channel is a regular polygon ring with parameters $n=8$, $r_i=20~\upmu$m, and $r_o=25~\upmu$m. Other parameters used for simulation are: simulation time steps of $\Delta T = 0.1$ seconds, MT diffusion coefficient $D = 2.0 \cdot 10^{-3}$ $\upmu$m$^2$/s, average speed of the MT $v_{\mathrm{avg}} = 0.5$ $\upmu$m/s,  and persistence length of the MT trajectory $L_p = 111$ $\upmu$m. We also assume the size of the information particles is 1$\upmu$m, the average length of the MTs is 10$\upmu$m, and each MT can load up to 5 information particles in one trip from the transmission zone to the receiver zone. These parameters are all selected based on experimental observations of DNA covered MTs moving over a kinesin covered substrate, and the reader is referred to \cite{far12NanoBio} for more details about the simulation environment. In the rest of the paper we will always use these parameters unless it is specified otherwise.

In Figure \ref{fig:SimEnv} the MT initially starts on top of the channel. It will then move towards the transmission zone (blue/cyan grid) along the left wall of the channel. The MT loads the first 5 information particles on its way and continues to the receiver zone where it unloads the loaded information particles. Notice that although the MT enters grid squares that contain information particles, after 5 loaded particles, it does not load anything else because the maximum load capacity is reached.

\section{Information Transmission Model}
\label{sec:chanCap}

In molecular communication, messages can be encoded into information carrying particles using many different schemes. Information can be encoded into the number, concentration, or type of the particles released. One key observation in molecular communication is that, regardless of the encoding scheme, information is always transmitted through mass transfer (i.e. transfer of particles). Therefore, in this section we present a general information transmission model for mass transfer, which is independent of the information encoding scheme.

Let $\mathcal X=\left\{ 0,1, 2, \cdots , x_{\text{max}}\right\}$ be the set of possible particles that could be released by the transmitter, where $x_{\text{max}}$ is the maximum number of particles the transmitter can release.  Let $X \in \mathcal X$ be the number of information particles released into the medium by the transmitter. We define {\em time per channel use} (TPCU) as a predefined amount of time $T$ representing the duration of a single transmission session. Because of the random motion of the MT filaments, given this time limit, information might not be perfectly conveyed to the receiver. For example, it is possible that some of the particles will not arrive at the receiver after $T$ has elapsed and therefore there is some information loss. This effect is similar in nature to the noise introduced by conventional electronic or electromagnetic channels. 

Let $Y^{(T)} \in \mathcal X$ represent the number of information particles that arrive at the destination after TPCU duration $T$. Because the receiver does not have any prior knowledge of the number of particles released by the transmitter, at the receiver $X$ is a discrete random variable given by probability mass function (PMF)  $f_X(x) $. Similarly, given $X$ particles were released by the transmitter, and the TPCU duration $T$, $Y^{(T)}$ is a discrete random variable given by conditional PMF $f_{Y^{(T)}\mid X} (y^{(T)} \mid x)$. This conditional PMF is very important because it characterizes the channel completely. It can be used to calculate important channel parameters such as {\em channel capacity} \cite{cover-book}, the maximum rate at which any communication system can {\em reliably transmit information} over a noisy channel.

Assuming that the transmitter and receiver are perfect (i.e. they release and receive the information particles perfectly), and that the information particles are not lost in the environment, the only factor that effects the PMF $f_{Y^{(T)}\mid X} (y^{(T)} \mid x)$ is the random motion of the MTs. Although this assumption may not hold in practice, by assuming perfect transmitter and receiver, we only focus on the effects of channel-shape on the information rate. Moreover, for some applications such as pure mass transport in lab-on-chip devices, we are only concerned with mass transport as opposed to a communication system. The PMF $f_{Y^{(T)}\mid X} (y^{(T)} \mid x)$ could be used as an important performance measure in both cases. For a given TPCU duration $T$, as the number of MT trips between the transmitter and the receiver increases, the number of particles that could potentially be transported increases. This results in higher mass transfer and potentially higher achievable information rates. Based on this observation we develop an optimization model which can be used to find the optimal channel shape.

\section{Channel Shape Optimization Model} 
\label{sec:model}
Let the set $\mathcal{G}$ be the set of all possible cross sectional geometric shapes the channel could have, where the cross sectional shape is the shape of the channel when viewed from the top. In this paper we fist assume this set contains all the rectangular, regular polygonal, and regular polygonal ring shapes and then extend the result to all two dimensional shapes. Our goal is to find the optimal shape in these sets, given system parameters such as TPCU, concentration of MTs, and average speed of the MTs. To setup this optimization problem, we model the effects of channel shape on the number of MT trips.

As discussed in the previous section, the number of MT {\em trips} during the TPCU interval has a direct effect on transportation of information particles. Without loss of generality let $g\in\mathcal{G}$ be the channel under consideration. Therefore, let $K^{(T)}(g)$ be the number of trips during the TPCU interval $T$ inside the channel shape $g$, where a {\em single MT trip} is defined as the movement of the MT from anywhere in the channel to the transmission zone and then the receiver zone. For example, a single MT trip is shown in Figure \ref{fig:SimEnv}. After the MT completes its first trip, subsequent trips are defined as the movement of the MT from the receiver zone to the transmission zone and back. During any trip, an MT can deliver 0 or more information particles (up to its maximum load capacity). Because the motion of the MT filaments are random in nature, $K^{(T)}(g)$ is a random variable. Therefore, to derive our optimization model we use the expected value of $K^{(T)}(g)$.

Let the random variable $K_s^{(T)}(g)$ be the number of trips for a single MT during the TPCU interval, $T$. Let $v_{\mathrm{avg}}$ be the average speed of the MTs, $P(g)$ be the perimeter of the channel shape $g\in\mathcal{G}$. A good estimate for the average number of MT trips, when a single MT is inside the channel is given by 
\begin{equation}
\label{eq:SingMTEst}
	E[K_s^{(T)}(g)] \approx \frac{v_{\mathrm{avg}}T}{P(g)}.
\end{equation}

\begin{figure}
	\begin{center}
	\includegraphics[width=3.4in]{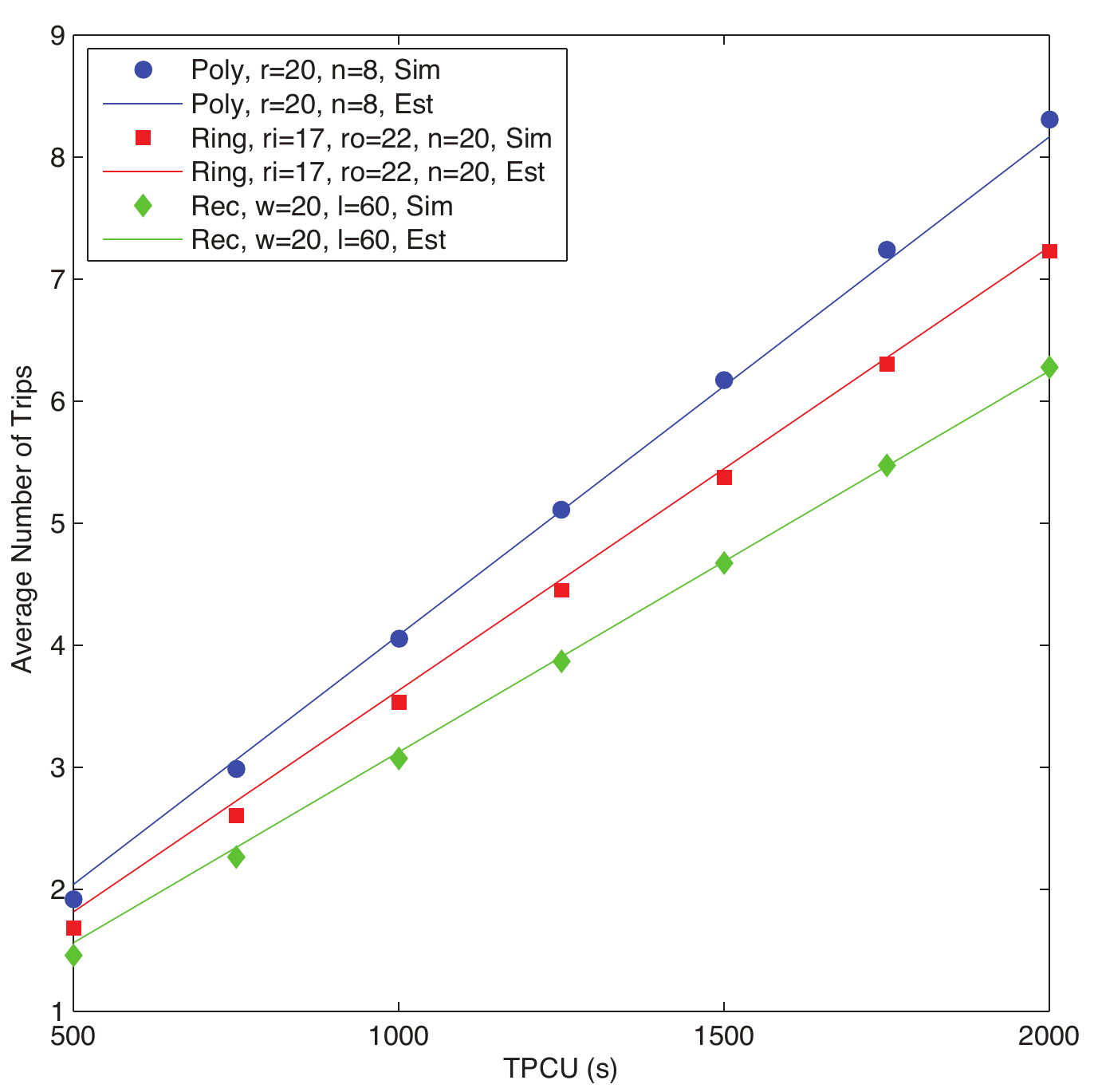}
	\end{center}
	\caption{\label{fig:avgTrip} Average number of trips approximation compared with Monte Carlo simulations.}
\end{figure}

This approximation is based on the observation that on average from each trip, the MT travels a distance equivalent to the perimeter of the channel. 
This assumption is verified using our Monte Carlo simulation environment. The results of the simulations are shown in Figure \ref{fig:avgTrip}. In this figure, we have considered a sample shape from each of our three different shape classes, and have shown that the approximations in Equation (\ref{eq:SingMTEst}) (solid lines) are a good estimate of the actual average number of MT trips (point plots).

In practice, there are typically more than one MT inside the channel. Furthermore, the number of MTs is not a constant, but is dependent on the volume of fluid in the channel: the DNA-covered MTs are prepared in chemical solutions, and therefore have a constant concentration inside the solution. Let $C$ be the concentration of MTs as number of MTs per unit volume, and $h$ be the height of the channel, and $A(g)$ be the cross sectional area of the channel with geometric shape $g\in\mathcal{G}$. Then the number of MTs inside the channel is given by
\begin{equation}
\label{eq:NumbMT}
	M(g) =  A(g) \times h \times C, 
\end{equation}
where $M(g)$ is the number of MTs inside channel $g$.

For each MT in the channel let the random variables $K_i^{(T)}(g)$ be the number of trips during $T$ seconds inside the channel with shape $g$, for $i=1,2,\cdots ,M$. Then, the total number of MT trips during $T$ seconds is given by
\begin{equation}
	K^{(T)}(g) = \sum_{i=1}^{M} K_i^{(T)}(g).
\end{equation}
The average number of MT trips during $T$ seconds is therefore calculated as
\begin{equation}
\label{eq:ExpTotal}
	E[K^{(T)}(g)] = \sum_{i=1}^{M} E[K_i^{(T)}(g)].
\end{equation}
We assume that the number of trips for individual MTs are independent and identically distributed, because they are chemically similar and don't interact with each other. Generally, this assumption has been found to be  valid in laboratory experiments \cite{nit10, hiy10LabChip}. Due to the identical distributions, the equation simplifies to
\begin{equation}
	E[K^{(T)}(g)] = \sum_{i=1}^{M} E[K_s^{(T)}(g)] = M(g) E[K_s^{(T)}(g)].
\end{equation}
Using the approximation shown in Equation (\ref{eq:SingMTEst}), and Equation (\ref{eq:NumbMT}), the total number of trips can be estimated as
\begin{equation}
\label{eq:MTtripEst}
	E[K^{(T)}(g)] \approx   T \times v_{\mathrm{avg}} \times C \times h \times \frac{A(g)}{P(g)},
\end{equation}
where $A(g)$ and $P(g)$ are the cross sectional area and perimeter of the channel with geometric shape $g \in \mathcal{G}$, respectively. The channel shape optimization problem can then be formulated as
\begin{equation}
\label{eq:OptiGen}
	\max_{g\in\mathcal{G}} E[K^{(T)}(g)].
\end{equation}
Assuming all the other parameters are constant except the cross sectional shape (including the height of the channel), the optimization problem becomes
\begin{equation}
\label{eq:OptiEq}
	\max_{g\in\mathcal{G}} \left[ \frac{A(g)}{P(g)} \right].
\end{equation}
Equation (\ref{eq:OptiEq}) is very interesting because it states that the optimal shape is the one with the largest area to perimeter ratio. However, there is an important constraint that must be satisfied. The perimeter of the channel must be small enough such that a single MT can complete at least a single trip. Therefore, we assume the perimeter must be such that its length can be traveled by the MT, on average, during the TPCU interval $T$. This constraint ensures that the perimeter is small enough such that on average enough information particles can be delivered during the given TPCU duration. Finally, the optimization problem can be written in its complete form as
\begin{align}
\label{eq:OptiFinal}
	\max_{g\in\mathcal{G}} \left[ \frac{A(g)}{P(g)} \right] \notag \\
	\text{subject to~} P(g) \leq  T  v_{\mathrm{avg}}.
\end{align}

\section{Optimal Shape Analysis}
\label{sec:OptiShapAna} 
Using the optimization formula derived in the previous section, we analyze the optimal channel design for each shape class individually, and then the optimal design over all.

\subsubsection{Rectangular Channels}
Let $\mathcal{G}_{\text{Rec}}$ be the set of all rectangular shapes. Because any rectangular shape can be characterized by the two parameters width $w$ and length $l$, the Equation (\ref{eq:OptiFinal}) becomes
\begin{align}
\label{eq:OptiRec}
	\max_{(w,l)\in\mathcal{G}_{\text{Rec}}} \left[ \frac{w \times l}{2w+2l} \right] \notag \\
	\text{subject to~} 2w+2l \leq T v_{\mathrm{avg}}.
\end{align}
Solving this optimization problem, the optimal channel design given both TPCU interval $T$ and average speed of the MTs $v_{\mathrm{avg}}$, is $w=l=0.25 T  v_{\mathrm{avg}}$. Therefore, for rectangular channels, the optimal channel shape is always the square shaped channel.

\subsubsection{Regular Polygonal Channels}
Let $\mathcal{G}_{\text{Poly}}$ be the set of all regular polygons. Regular polygons can be characterized by two parameters: the number of sides $n$, and the radius of the circumscribed circle $r$. Note that the set of all square shapes are also contained within the set $\mathcal{G}_{\text{Poly}}$ (i.e. when $n=4$). Using these parameters Equation (\ref{eq:OptiFinal}) becomes 
\begin{align}
\label{eq:OptiPoly1}
\max_{(n,r)\in\mathcal{G}_{\text{Poly}}} \left[ \frac{0.5 n r^2 \sin (2\pi /n)}{2 n r \sin(\pi /n)} \right] \notag \\
	\text{subject to~} 2 n r \sin(\pi /n) \leq T  v_{\mathrm{avg}}.
\end{align}
Simplifying this equation we get
\begin{align}
\label{eq:OptiPoly2}
\max_{(n,r)\in\mathcal{G}_{\text{Poly}}} 0.5 r  \cos(\pi /n) \notag \\
	\text{subject to~} n r \sin(\pi /n) \leq \frac{T  v_{\mathrm{avg}}}{2},
\end{align}
where we have used the fact that $\sin (2u) = 2\sin(u)\cos(u)$. Based on this equation the optimal channel is the one with $n = \infty$ and $r=T  v_{\mathrm{avg}}/2\pi$. Therefore, when regular polygons are considered the optimal shape is the circular-shaped channels. Because squares are also in the set of all regular polygons, we conclude that circular-shaped channels are also better than square-shaped channels.

\subsubsection{Regular Polygonal Ring Channels}
Let $\mathcal{G}_{\text{Ring}}$ be the set of all polygonal ring shapes. The elements of this set can be characterized using three parameters: the number sides $n$, the radius of the inner polygon's circumscribed circle $r_i$, and the radius of the outer polygon's circumscribed circle $r_o$. Note that $\mathcal{G}_{\text{Ring}}\supset \mathcal{G}_{\text{Poly}}$, which means that the set $\mathcal{G}_{\text{Ring}}$ contains all the regular polygonal shapes (this is achieved by setting $r_i=0$). Using these parameters the optimization problem becomes
\begin{align}
\label{eq:OptiRing1}
\max_{(n,r_i,r_o)\in\mathcal{G}_{\text{Ring}}} \left[ \frac{0.5 n (r_o^2-r_i^2) \sin (2\pi /n)}{2nr_o\sin(\pi /n)} \right] \notag \\
	\text{subject to~} 2nr_o\sin(\pi /n) \leq T v_{\mathrm{avg}}.
\end{align}
Simplifying this equation we get
\begin{align}
\label{eq:OptiRing2}
\max_{(n,r_i,r_o)\in\mathcal{G}_{\text{Ring}}} \left[  0.5 \left(\frac{r_o^2-r_i^2}{r_o}\right)  \cos(\pi /n) \right] \notag \\
	\text{subject to~} nr_o\sin(\pi /n) \leq \frac{T v_{\mathrm{avg}}}{2}.
\end{align}
Solving Equation \ref{eq:OptiRing2}, the optimal channel has parameters $n = \infty$, $r_o=T  v_{\mathrm{avg}}/2\pi$, and $r_i=0$. This means the optimal channel is the circular-shaped channel. This may seem surprising at first, since one would expect that a ring-shaped circular channel would be better than a solid circular-shaped channel at transporting information particles. However, because typically MTs follow the walls of the channel, and that the number of MTs in a channel is proportional to the volume of the channel \cite{cle03}, it becomes apparent that the solid circular channel would be better than a ring-shaped channel.

\subsubsection{Overall Optimal Channel Design} 
From the solution to the optimization formulas for each of the three different shape classes, it is apparent that the optimal channel shape is the circular-shaped channel when the set of geometric shapes $\mathcal{G}_{\text{RPR}}=\mathcal{G}_{\text{Rec}}\cup\mathcal{G}_{\text{Poly}}\cup\mathcal{G}_{\text{Ring}}$ is considered. Moreover, based on Equation (\ref{eq:OptiFinal}) the optimal channel shape must have the largest area to perimeter ratio. Therefore, if the perimeter of the channel is fixed such that the constraint in the optimization is satisfied with an equality, the optimal channel shape would be a circle for all two dimensional geometric shapes.  

\section{Results and Discussions} 
\label{sec:results} 
To verify our optimization formulas and their results, we rely on Monte Carlo simulation. For these simulations we use the same parameters discussed in Section \ref{sec:ChanSim}. Moreover, we assume the height of the channel is always a constant $h=10$ $\upmu$m regardless of the cross sectional shape of the channel. The concentration of the MTs is also assumed to be $C = 0.001$ MT/fL unless specified otherwise. The number of MT in the channel is always assumed to be
\begin{equation}
	MT = \lfloor A(g)hC \rfloor,
\end{equation}
unless specified otherwise.

For the performance measure we rely on {\em channel capacity} \cite{cover-book}, which is the maximum rate at which information can be transmitted reliably. Channel capacity is given by
\begin{equation}
	C = \max_{f_X(x)} I(X;Y^{(T)}),
\end{equation}
where $I(X;Y^{(T)})$ is the mutual information between $X$ and $Y^{(T)}$. Mutual information is defined as 
\begin{equation}
	I(X;Y^{(T)}) = E\left[ \log_2 \frac{f_{Y^{(T)}|X}(y^{(T)}|x)}{\sum_x f_{Y^{(T)}|X}(y^{(T)}|x)f_X(x)} \right],
\end{equation}
where $f_{Y^{(T)}|X}(y^{(T)}|x)$ represents the probability of receiving symbol $y^{(T)}$ at the destination during TPCU interval $T$, given that symbol $x$ was transmitted by the source; $f_X(x)$ represents the probability of transmitting symbol $x$; and $E[\cdot]$ represents expectation. Using Monte Carlo simulations the PMF $f_{Y^{(T)}|X}(y^{(T)}|x)$ is estimated for a given channel shape. For more details about the Monte Carlo simulations for estimating this PMF please see \cite{far12NanoBio}, and for analytical estimation of this  PMF using Markov chain models please refer to \cite{far14TSP}. In this work, we use Monte Carlo simulations since they are more accurate in multi-MT environments. Using the estimated PMF $f_{Y^{(T)}|X}(y^{(T)}|x)$, the maximizing PMF $f_X(x)$ and hence the channel capacity is calculated using Blahut-Arimoto algorithm \cite{bla1972, ari72}. 

We now consider each of three shape classes discussed in the previous sections, and use the Equations (\ref{eq:OptiRec}), (\ref{eq:OptiPoly2}), and (\ref{eq:OptiRing2}) to find the optimal channel dimensions for each case. These results are then compared to the channel capacities obtained from Monte Carlo simulations.

\subsection{Rectangular Channels}
\begin{figure*}
   \subfloat[\label{fig:SimRecOpti160}]{
   										\includegraphics[width=3.4in]{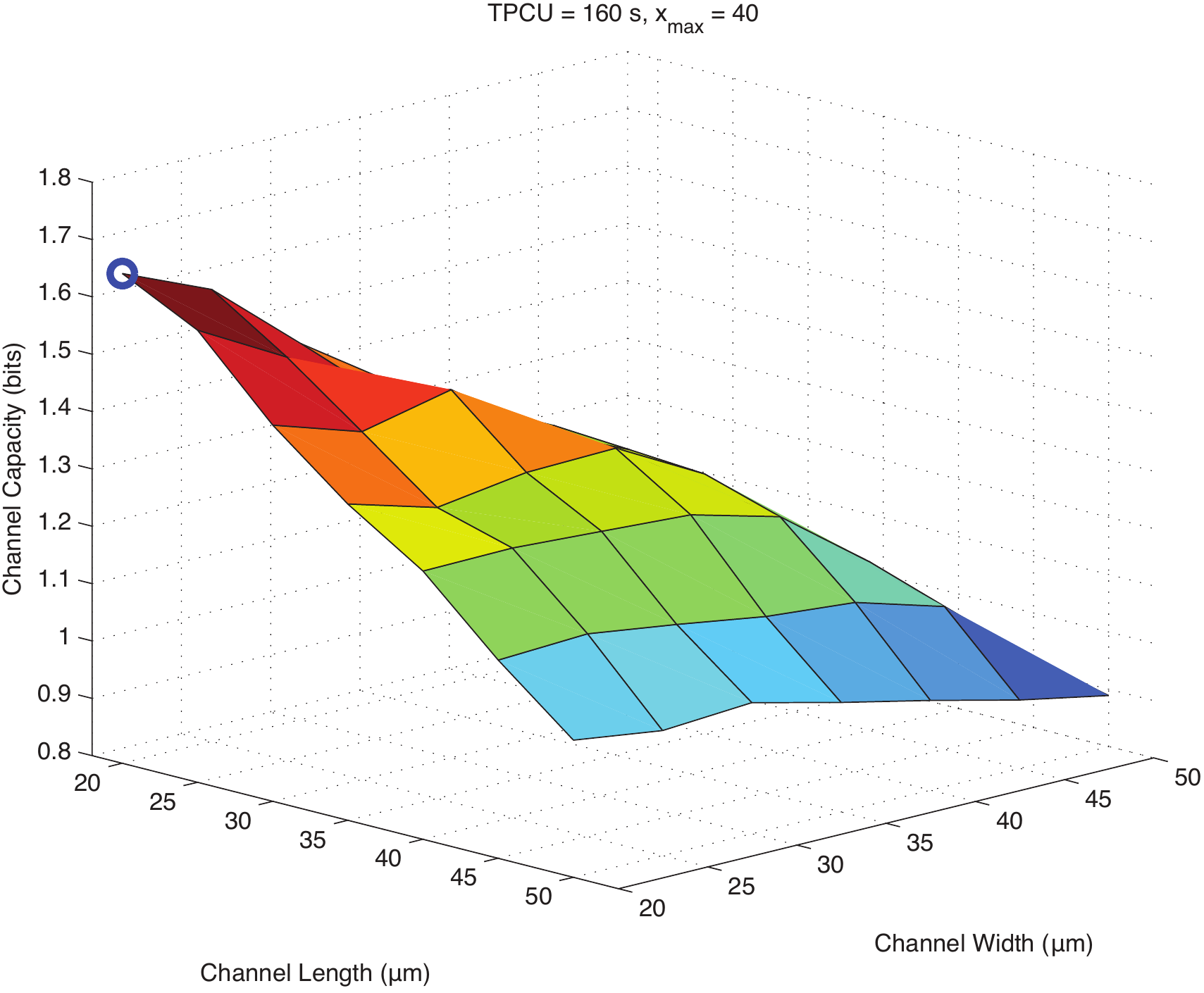}
   									   }
   \subfloat[\label{fig:SimRecOpti240}]{
   										\includegraphics[width=3.4in]{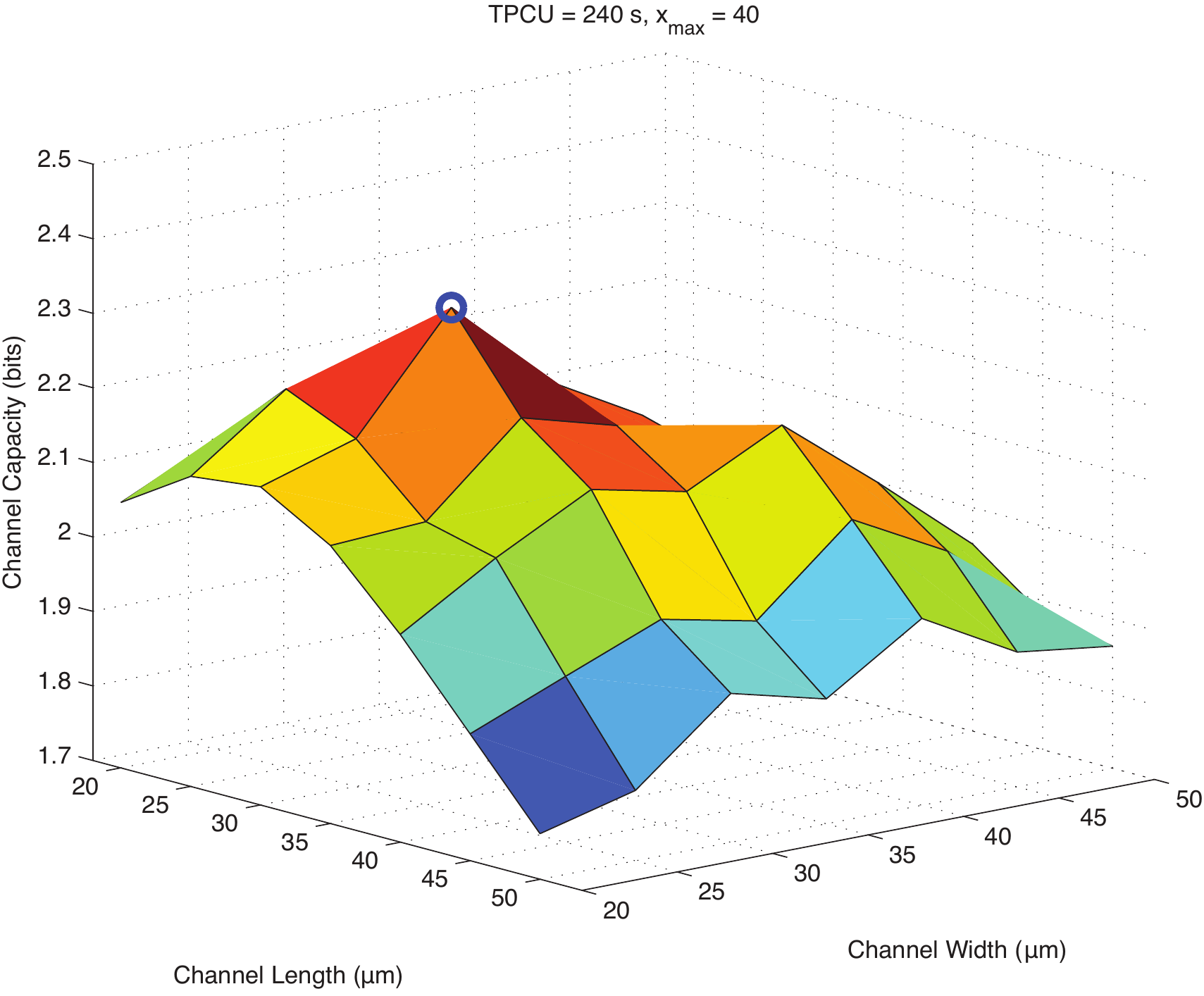}
   										}
   \caption{\label{fig:SimRecOpti} Channel capacities of different rectangular channel shapes. The channel with the highest capacity is shown using the blue dot. For TPCU of 160 seconds the optimal channel is 20 $\upmu$m $\times$ 20 $\upmu$m (a), and for TPCU value of 240 seconds the optimal channel dimensions is 30 $\upmu$m $\times$ 30 $\upmu$m (b).}
\end{figure*}
First, we consider rectangular channels. We consider two values of TPCU: $T=160$ seconds, and $T=240$ seconds. Using the solution to Equation (\ref{eq:OptiRec}), the optimal channels are square channels with dimensions $w=l=0.25\times 160\times 0.5=20$ $\upmu$m and $w=l=0.25\times 240\times 0.5 = 30$ $\upmu$m for the  $T=160$ and  $T=240$, respectively. To verify this result, we simulate all the rectangular channels with widths and lengths raging from $20$ $\upmu$m to $50$ $\upmu$m in $5$ $\upmu$m steps. We assume $x_{\text{max}}$ (i.e. the maximum number of particles the transmitter can release) is 40. The channel capacity is then calculated based on the simulations. Figure \ref{fig:SimRecOpti} shows the results for $T=160$ seconds (a), and $T=240$ seconds (b). As can be seen the channels with the largest channel capacity are the 20 $\upmu$m $\times$ 20 $\upmu$m, and 30 $\upmu$m $\times$ 30 $\upmu$m, respectively. These results are in perfect agreement with the results obtained from our optimization model.

\subsection{Regular Polygonal Channels}
\begin{figure}
	\begin{center}
	\includegraphics[width=3.4in]{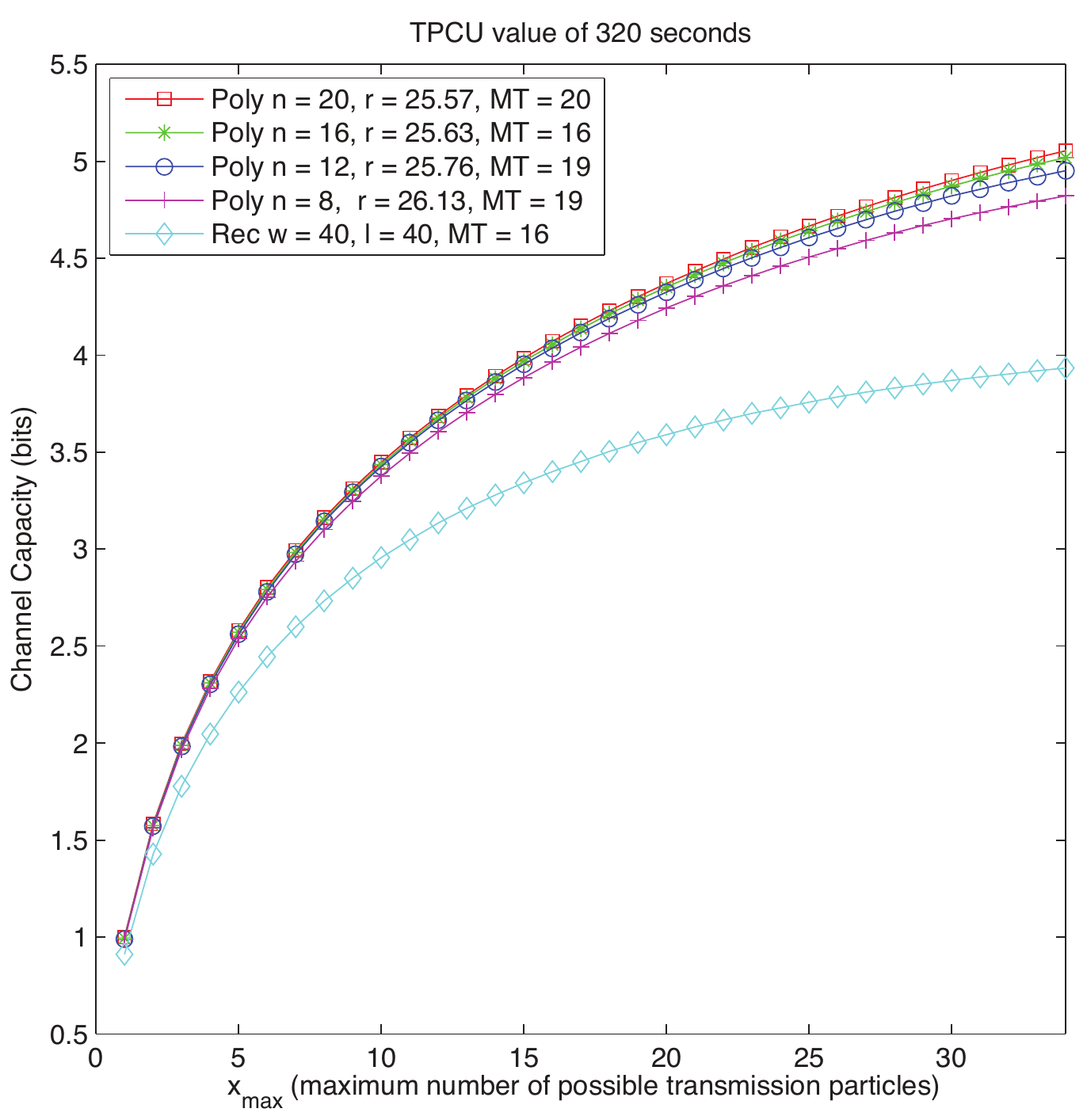}
	\end{center}
	\caption{\label{fig:SamePeri} Channel capacity of regular polygonal channels with constant perimeter of 160 $\upmu$m versus maximum number of particles that can be released at the transmitter.}
\end{figure}
In Section \ref{sec:OptiShapAna} we showed that for regular polygonal channels the optimal shape is a circular-shaped channel. To verify this result, we first consider a constant TPCU value of 320 seconds. From Equation (\ref{eq:OptiFinal}), the maximum channel perimeter that satisfies the constraint is $160$ $\upmu$m. Therefore, we assume that the channel perimeter is fixed at this value and show that the capacity increases a the channel shape become more circular by increasing the number of sides $n$. In particular we consider the following channels: square channel of length 40 $\upmu$m (equivalent to a polygon with $n=4$), and regular polygonal channels with parameters ($n=8$, $r=26.13$ $\upmu$m), ($n=12$, $r=25.76$ $\upmu$m), ($n=16$, $r=25.63$ $\upmu$m), and ($n=20$, $r=25.57$ $\upmu$m). Based on our optimization formula we expect the 20-sided channel to be the optimal among all these channels since it is more circular. Figure \ref{fig:SamePeri} shows the channel capacity for each of these channels obtained through Monte Carlo simulations versus the maximum number of particles the transmitter can release $x_{\text{max}}$. As can be seen, the optimal channel is indeed the 20-sided regular polygon. Moreover, from the obtained pattern it is obvious that as the number of sides increase the channel capacity increases. This supports the results obtained from our optimization formula that circular channels are optimal.

\begin{figure}
	\begin{center}
	\includegraphics[width=3.4in]{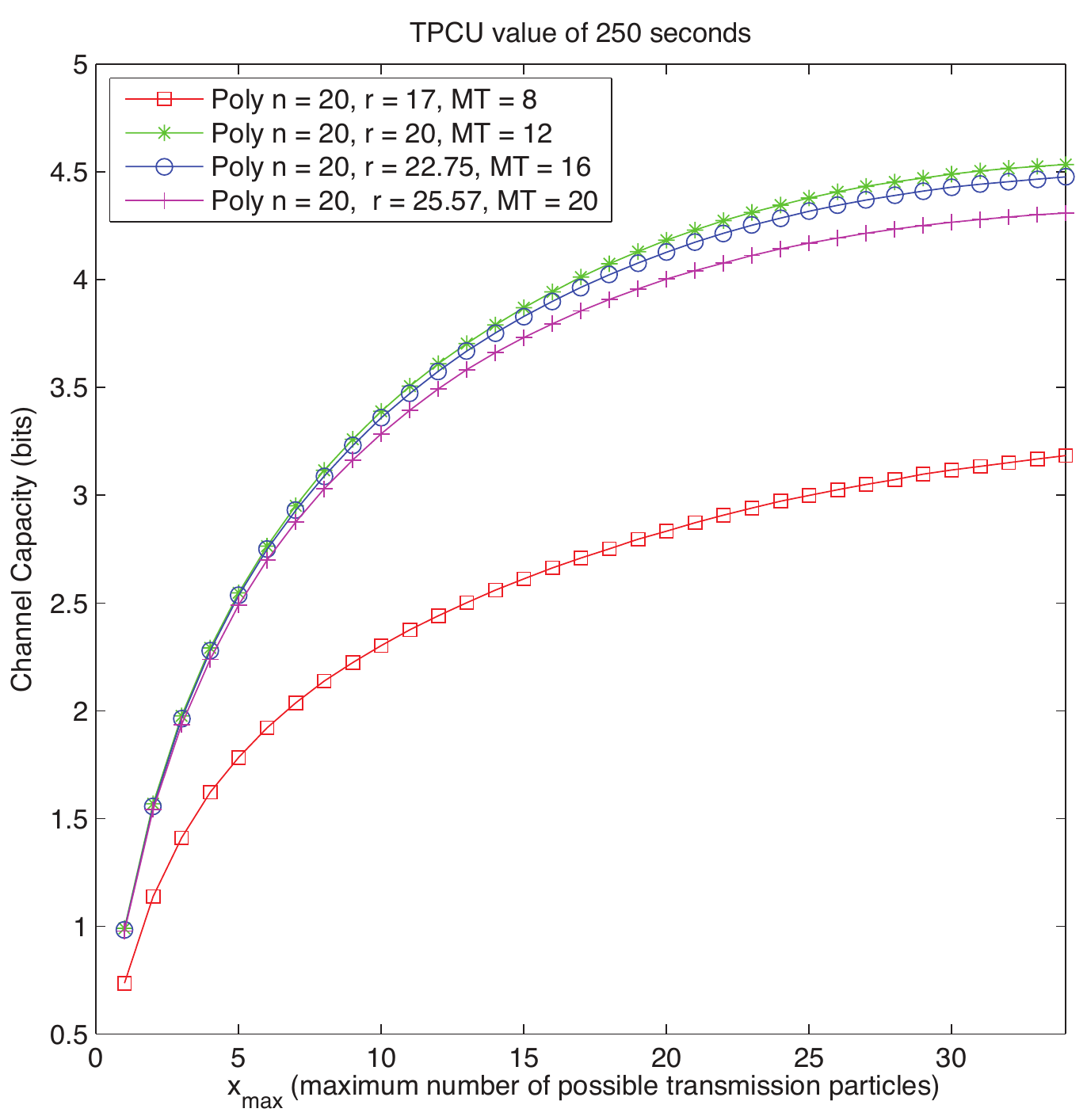}
	\end{center}
	\caption{\label{fig:SameSides} Channel capacity of four circular-shaped channels versus maximum number of particles that can be released at the transmitter.}
\end{figure}
To evaluate the performance of Equation (\ref{eq:OptiPoly2}) we consider the TPCU interval of $T=250$ seconds. The solution of this optimization problem is a channel with parameters $n=\infty$ and $r\approx20$ $\upmu$m. Because in our simulation environment the number of sides must be a finite number, we use the value of $n=20$ to simulate a perfect circle. We then simulate four different regular polygonal channels with parameters: ($n=20$, $r=17$ $\upmu$m), ($n=20$, $r=20$ $\upmu$m), ($n=20$, $r=22.75$ $\upmu$m), and ($n=20$, $r=25.57$ $\upmu$m). Figure \ref{fig:SameSides} shows the channel capacity of each channel versus the maximum number of particles the transmitter could release. As can be seen in the figure, the channel with the radius of 20 $\upmu$m achieves the highest capacity among the four channels. This result is in agreement with the optimal solution derived from our optimization formula.

\subsection{Regular Polygonal Ring Channels}

\begin{figure}
	\begin{center}
	\includegraphics[width=3.4in]{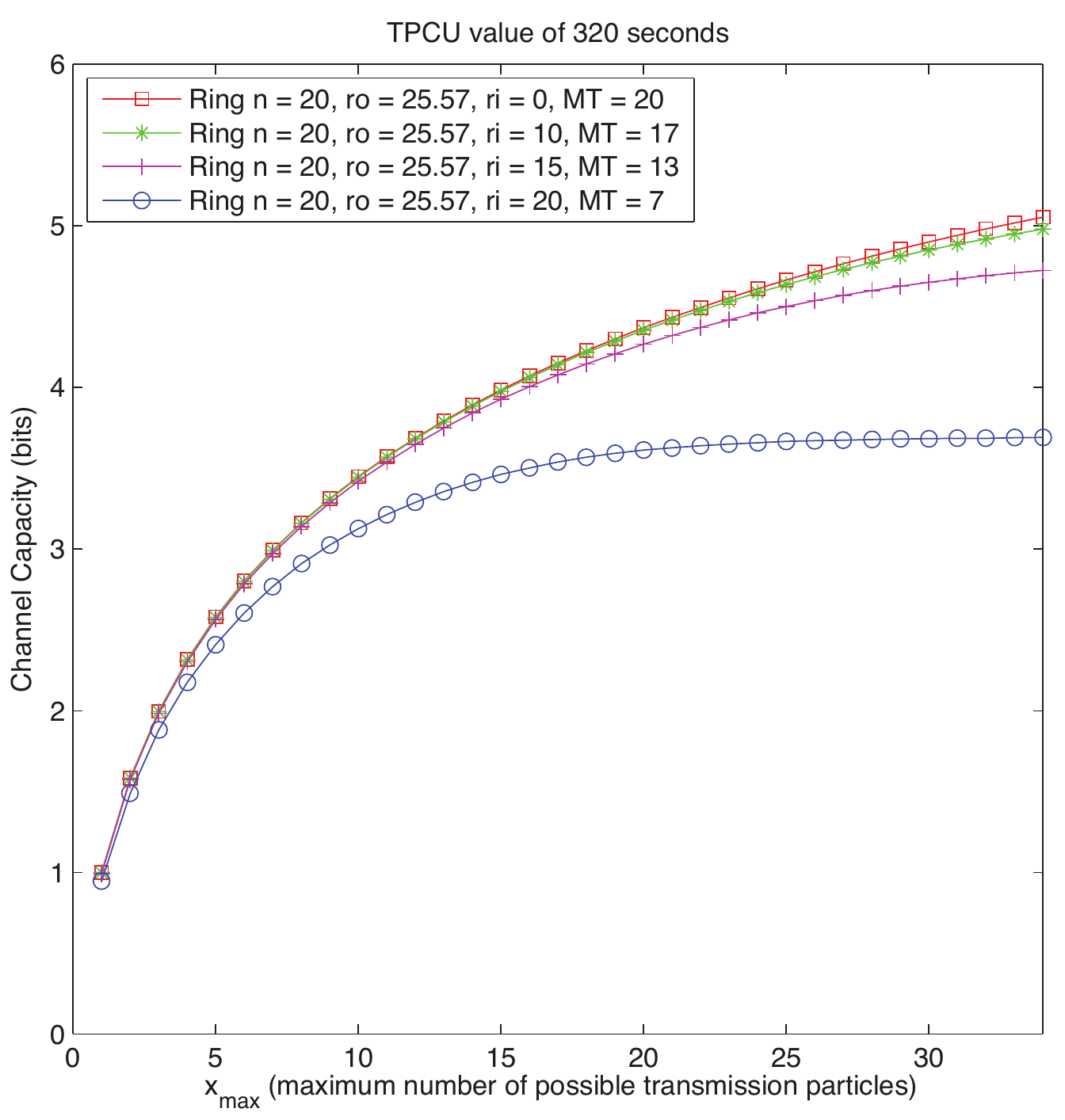}
	\end{center}
	\caption{\label{fig:RingChan} Channel capacity of ring-shaped channels versus maximum number of particles that can be released at the transmitter.}
\end{figure}
In this section, we validate our optimization formula for polygonal ring-shaped channels presented in Equation (\ref{eq:OptiRing2}) through simulations. We assume the TPCU interval is fixed at 320 seconds. Using the solution to Equation (\ref{eq:OptiRing2}), the optimal channel dimensions are $n=\infty$, $r_i=0$, and $r_o \approx 25.57$ $\upmu$m. In Figure \ref{fig:SamePeri} we have already shown that the solid circular-shaped channel is optimal among regular polygons. Therefore, we only consider ring-shaped channels for validation. In particular, we consider the following ring-shaped channels: ($n=20$, $r_o=25.57$ $\upmu$m, $r_i=0$ $\upmu$m), ($n=20$, $r_o=25.57$ $\upmu$m, $r_i=10$ $\upmu$m), ($n=20$, $r_o=25.57$ $\upmu$m, $r_i=15$ $\upmu$m), and ($n=20$, $r_o=25.57$ $\upmu$m, $r_i=20$ $\upmu$m). Figure \ref{fig:RingChan} shows the channel capacity of these channels versus the maximum number of particles the transmitter could release. We can see from the figure that a solid circular channel achieves the highest capacity as predicted by our optimization formula.

Based on all these results, we can see that a solid circular-shaped channel is the optimal shape that maximizes information rate, when stationary kinesin with mobile MT filaments are used for transportation in molecular communication systems. Moreover, as the number of sides increase and the channel becomes closer to circular, the performance gains of adding more sides decreases. Based on the obtained results, there is negligible performance difference between channels with the number of sides greater than 20. Therefore, these channels are sufficient representation of perfectly circular-shaped channels.

\section{Conclusion} 
\label{sec:conclusion} 
In this work we considered the problem of finding the optimal channel shape for on-chip molecular communication based on kinesin driven microtubule (MT) motility. To setup our optimization model, we first derived a mathematical model relating the average number of MT trips to the shape of the channel. Using this model we then presented the three optimization formulas and their respective solutions for three different classes of shapes: the rectangular, regular polygonal, and regular polygonal rings. We showed that the optimal solution is the square channel for the rectangular channel shapes, and circular shape for the regular polygonal and ring-shaped channels. Finally, we showed that when all classes of 2 dimensional shaped are considered the circular-shaped channel tend to be the one that achieves the highest information rate. Moreover, using our optimization formula we showed that the dimensions of the optimal channel can be reliably estimated.

\renewcommand{\baselinestretch}{1}
\normalsize

\bibliography{MolCom}{}
\bibliographystyle{ieeetr} 

\end{document}